\documentclass[12pt]{amsart}

\usepackage{amssymb,amscd}
\usepackage{verbatim}
\usepackage{graphicx}

\newtheorem*{Whitney towers}{Theorem~\ref{Whitney towers}}
\newtheorem*{h-towers}{Theorems ~\ref{half} \& \ref{$(n)$-solvable}}

\newtheorem*{surgery curves}{Theorem~\ref{surgery curves}}
\newtheorem*{cg=0}{Theorem~\ref{vanish}}

\theoremstyle{definition}

\numberwithin{equation}{section}
\numberwithin{figure}{section}

\newcommand{\nl}{\newline}
\newcommand{\Z}{\mathbb{Z}}
\newcommand{\N}{\mathbb{N}}

\newcommand{\Q}{\mathbb{Q}}

\title{
`Topological parallel world' 
constructed by modification of space-time along observables
}
\author{Eiji Ogasa}
\address{
High Energy Physics Theory Group\\
Department of Physics\\
Ochanomizu University\\ 
Bunkyo-ku\\
Tokyo 112-8610\\ 
JAPAN\\}
\email{ogasa@phys.ocha.ac.jp}

\thanks{
\nl{\bf PACS nos.} 11-25w, 11-25Uv.}

\begin{document}
\begin{abstract} 
We introduce a new concept, 
`(topological) (vacuum) parallel world, ' 
which is a new tool to research submanifolds.  
Roughly speaking, 
`Observables in (T)QFT' is equal to 
`a (topological) modification of space-time.' 
In other words, we give a new interpretation of observables. 
\end{abstract} 
\maketitle

\section{Introduction: An Example} 

In this paper 
we introduce a new concept, 
`(topological) (vacuum) parallel world, ' 
which is a new tool to research submanifolds.  

In this section  
we give an example of topological vacuum parallel worlds. 
In \S2 we give definition.  
In \S3 we give one more example and problems. 
In \S4 we have discussion.

\vskip3mm
\noindent{\bf Example 1. 
The property 
`whether or not the Alexander polynomial 
for an arbitrary 1-knot in $S^3$ is trivial'  
and its parallel world. }  

Let $K$ be a 1-knot in $S^3$. 
Let $\Delta_K(t)$ be the Alexander polynomial of $K$.  
See \cite{Rolfsen} for 1-knots and so on.

\noindent
{\bf One way to define the above property:}
\cite{Bar-NatanGaroufalidis} 
\cite{Rozansky} 
\cite{Vaintrob96} 
\cite{Vaintrob97} 
etc. say:
 $\Delta_K(t)$ can be represented by an information of a path integral 
 \newline 
 $Z(S^3, K)=\int_{\mathcal G} {\mathcal D}A 
{\mathrm {exp}}
(\frac{ik}{4\pi}\int_M{\rm Tr}\{{\mathcal {CS}}(A)\})
{\mathcal O}(K),$ 
where ${\mathcal G}$ is a space of sections on a bundle of $S^3$ 
and ${\mathcal O}(K)$ is an observable defined by using Wilson line 
(see the following note).

\noindent
{\bf Note.} 
Let $M$ be a closed oriented 3-manifold. 
Let $L\subset M$ 
be a submanifold diffeomorphic to some copies of $S^1$. 
\cite{WittenQFTJones} says 
a path integral 
$$Z_k(M, L)=\int_{\mathcal G} {\mathcal D}A 
{\mathrm {exp}}
(\frac{ik}{4\pi}\int_M{\rm Tr}\{{\mathcal {CS}}(A)\})
{\rm Wil}(L),$$
\newline 
where $A$ is a 1-form on $M$ associated with a flat $G$ bundle, 
where ${\mathcal {CS}}(A)$ is the Chern-Simon 3-form, 
where $G$ is a compact Lie group, 
where ${\mathcal G}$ is a space of sections on the bundle 
and 
where $k\in\Z$, 
\newline
essentially represents the Jones polynomial of $L$ 
if $M\cong S^3$ and $G=SU(2)$.  
Furthermore, 
if $L$ is the empty set, $Z_k(M)$ essentially represents 
Witten's quantum invariant of $M$.

\noindent
{\bf Another way to define the above property:} 
Suppose that an oriented closed 3-manifold $B(K,n)$ is
the branched cyclic covering space along $K$. 
(See \cite{Rolfsen}.)

\cite{DijkgraafWitten}   
says:  
the above $Z_k(M)$ is defined for 
if $G$ is a discrete Lie group. 
If we take an appropriate Lagrangean corresponding to the above,  
call it $Z(M)$, 
(which does not have an parameter $k$; see \cite{DijkgraafWitten}). 
Then $Z(M)$ is 
$\frac{|{\mathrm {Hom}}(\pi_1(M), G)|}{|G|}$, 
where $|\hskip3mm|$ is the order of a group.

Let $G=\Z_p$ for any $p\in\N-\{1\}$. 
Let $n\in\N-\{1\}$.  
Let 
$Z(B(K,n))_p=Z(B(K,n))$ if $G=\Z_p$. 
Then 
$Z(B(K,n))_p$ 
=$\frac{|{\mathrm {Hom}}(\pi_1(B(K,n)),\Z_p)|}{p}$
=$\frac{|{\mathrm {Hom}}(H_1(B(K,n);\Z),\Z_p)|}{p}$. 

If $Z(B(K,n))_p=\frac{1}{p}$ for all pair $(p,n)$, 
then $H_1(B(K,n);\Z)\cong0$  for all $n$. 
 \cite{Gordon} 
 \cite{Riley} 
 say that,  $\Delta_K(t)=1$ holds 
if and only if $H_1(B(K,n);\Z)\cong0$ for all $n$. 
Therefore we have: 
whether $\Delta_K(t)=1$ or not 
can be represented by another path integral.

Therefore we have: 
whether $\Delta_K(t)=1$ or not 
can be represented by two ways of path integrals.

\vskip3mm
Roughly speaking, 
the former way uses a path integral of a Lagrangean and an observable. 
The latter way uses a modification of space-time along an obserbable 
and 
a path integral of a Lagrangean, where this path integral does not include 
an observable. 
More roughly speaking, 
modification of space-time is `equal' to an observable in QFT.

After our definition in \S2,  
We say:  
$\{B(K,n)\}$ is 
a {\it parallel world} for all knots $K$ 
associated with 
whether  or not the Alexander polynomial is trivial. 
The operation to make the branched cyclic covering spaces 
is 
{\it modification of space-time} to make 
{\it  the parallel world} $\{B(K,n)\}$. 
See \S2 for detail. 

\noindent
Note: A parallel world is a set of manifolds in a case 
and is a single manifold in another case.

\section{Topological vacuum parallel world}

In this section 
we introduce `(topological) (vacuum) parallel world.' 

\vskip3mm
Let $X$ be a smooth $x$-dimensional manifold. 
Let $Y$ be a smooth $y$-dimensional manifold. 
Let $y\leqq x$. 
Let $K\subset X$ be a smooth submanifold of $X$ 
which is diffeomorphic to $Y$.

In this paper we research submanifolds.  For example, we consider 
whether or not a submanifold $K\subset X$ is equivalent to 
another submanifold $K'\subset X$. 
See 
\cite{LevineOrr}
\cite{MilnorStasheff} 
\cite{Rolfsen} 
for manifolds, submanifolds and so on.

\vskip3mm
\noindent
{\bf One way to define an invariant for submanifolds:}
A path integral 

$$Z(X, K)=\int_{\mathcal G} {\mathcal D}A 
{\mathrm {exp}}({\mathcal L}){\mathcal O}(K),$$

\noindent 
where ${\mathcal G}$ is a space of sections on $X$,  
where ${\mathcal L}$ is a Lagrangean,   
and where ${\mathcal O}(K)$ is an observable associated with $K$,  
 gives an invariant of submanifolds $K\subset X$. 
If, for some ${\mathcal L}$ and ${\mathcal O}(K)$, 
we have $Z(X, K)\neq Z(X, K')$, then 
the submanifold $K\subset X$ is not equivalent to 
the submanifold  $K'\subset X$. 

\vskip3mm 
\noindent 
{\bf Another way to define an invariant for submanifolds:} 
Take a topological modification $\mathcal E$ 
which makes $X$ into a new manifold $M_K$ 
by an operation associated with $K$.
Examples: surgery along knots, branched covering spaces along knots, 
$X-K$ etc. See \S1,3. 

A path integral 
$$Z(M_K)=\int_{\mathcal F} {\mathcal D}A 
{\mathrm {exp}}({\mathcal L}),$$

\noindent 
where ${\mathcal F}$ is a space of sections on $M_K$,  
 gives an invariant of submanifolds $K\subset X$. 
For some ${\mathcal L}$ and ${\mathcal E}$, 
we have $Z(M_K)\neq Z(M_{K'})$. 
It implies that 
the submanifold $K\subset X$ is not equivalent to 
the submanifold  $K'\subset X$.

\vskip3mm
\noindent
{\bf Definition.}
Fix $X$, $Y$ as above. 
If we have $Z(M_K)=Z(X, K)$ for any submanifold $K\subset X$,  
then we say 
that 
$M_K$ is a {\it {topological vacuum parallel world}} 
of 
the submanifold $K\subset X$ associated with $Z(X, K)$ 
and 
that the topological modification ${\mathcal E}$  makes 
a 
{\it {topological vacuum parallel world}} 
of submanifolds $K\subset X$ 
associated with $Z(X, K)$.

If some information of $Z(M_K)$ coincides with 
some information of $Z(X, K)$, then we also say that 
the topological modification ${\mathcal E}$  makes 
a {\it topological vacuum parallel world} for submanifolds $K\subset X$ 
associated with the information of $Z(X, K)$.

\vskip3mm

Roughly speaking, the following two are equivalent. 

\noindent 
(1) a path integral of a Lagrangean and an observable ${\mathcal O}(K)$.  

\noindent 
(2) a path integral of a Lagrangean 
associated with $M_K$  
and 
a topological modification ${\mathcal E}$ associated with $K$. 

\noindent 
In other words, more roughly speaking, 
`Observables in (T)QFT' is equal to 
`a (topological) modification of space-time.' 
We give a new interpretation of observables. 

\noindent 
Note: we call both  $K$ and ${\mathcal O}(K)$ observable.

\vskip3mm
The above definition is not so strict. 
One can change it a little for their purpose. 
For example, changing 
${\mathcal L}$ into $\frac{ik}{4\pi}{\mathcal L}$.
A parallel world is a single manifold in a case 
and is a set of manifolds in another case.  
We may discuss complex manifold case, symplectic ones and so on.

\section{Problems}

\noindent
{\bf Example 2.}
The Jones polynomial for $(2,\pm(5l+3))$-torus knots and  
 its parallel world($l\in\Z$ and $5l+3$ is an odd number).

Let $K$ be a $(2,\pm(5l+3))$-torus knot. 
Let  $V_K(t)$ be the Jones polynomial of $K$. 

One way to define the above one: 
As in Note in Example 1, 
the Jones polynomial is defined by a path integral.

Another way to define the above one: 
Let $S(K,m)$ be a closed oriented 3-manifold which is made from $S^3$ 
by a surgery along $K$ with framing $m$ (see \cite{Kirby}). 
Then 
$S(K,1)$ is a homology 3-sphere. 
Let $\lambda=\lambda(S(K,1))$  be the Casson invariant of $S(K,1)$. 
Consider an equation  
 $\lambda=\Sigma^n_{i=1}(-1)^{n+i}i^2$ on $n$. 
Then there is one and only one solution which is a nonnegative integer, 
call it $n$. 
Take 
\noindent
$f(t)=t^{n}(1-t^2)^{-1}
(1-t^{3}-t^{2n+2}+t^{2n+3})$ 
and  
$g(t)=f(\frac{1}{t})$. 

\noindent
By using the quantum invariant 
$\tau_5(S(K,m))$ $(m\in\Z)$, 
we can know  
$V_K(e^\frac{2\pi i}{5})$
(see Theorem (8.14) in \cite{KirbyMelvin}). 
In our case, 
only one of the polynomials $f(t)$ and $g(t)$ satisfies 
the condition that the value at $t=e^\frac{2\pi i}{5}$ 
is equal to the value $V_K(e^\frac{2\pi i}{5})$ 
(see  Th. 14.13 in \cite{Lickorishbook}), call it $h(t)$.
$V_K(t)$ is the polynomial $h(t)$.


\noindent
\cite{Murakami} says 
the Casson invariant of any homology 3-sphere 
is determined by the all quantum invariants of the homology 3-sphere. 
It means  $\lambda(S(K,1))$ can be represented by a path integral and,  
furthermore, 
that the Jones polynomial of the above $K$ 
can be represented by another path integral.  
$\{S(K,m)\}$ is a parallel world in this case.

\noindent 
Note: The gauge groups are same 
in both path integrals 
representing the Jones polynomial of these examples. 
Each of them is $SU(2)$.

\vskip3mm
By using 
\cite{AxelrodSinger} 
\cite{DijkgraafWitten}   
\cite{KirbyMelvin} 
\cite{Lickorish91} 
\cite{ReshetikhinTuraev} 
\cite{RozanskyWitten}
\cite{WittenQFTJones}
and their generalizations in many other papers, 
we can make such many examples of 
topological parallel worlds.  
For example, the value of the Jones polynomial and 
that of the Alexander polynomial at a value can be obtained  from 
the quantum invariants. 
Another example: the Alexander polynomial of torus knots can be obtained 
from $Z(B(K,n))_p$ for all $(p,n)$.

\vskip3mm
We give two problems. 

\noindent 
(1) Let $\tau_r$ be the $r$-th quantum invariant (see \cite{KirbyMelvin} etc). 
Suppose that  $\tau_r(B(K,n))=\tau_r(B(K',n))$  for any $n\in\N-\{1\}$ 
and for any $r$. 
Suppose that 
$\tau_r(S(K,n))=\tau_r(S(K',n))$  
for any $n\in\Q$ and for any $r$. 
Then do we have $V_K(t)=V_{K'}(t)$? 
(Note $S(K,n)$ can be defined for any rational number; see \cite{Rolfsen}).

\noindent 
(2) 
If $Z(B(K,n))_p=Z(B(K',n))_p$ holds, 
then 
 do we have $\Delta_K(t)=\Delta_{K'}(t)$? 

\noindent 
If $|H_1(B(K,n),\Z)|=|H_1(B(K',n),\Z)|$@for@any@$n$,  
then do we have $\Delta_K(t)=\Delta_{K'}(t)$? 


\section{Discussion}
It would be better 
that we say 
that the concept of parallel worlds for observables 
is a principle to find a new model of TQFT 
than 
that we regard the existence of the above examples as theorems. 
Because we can make a parallel world for any observable in a trivial way 
similar to the following way: 
Consider the Jones polynomial $V_K(t)=\Sigma_i a_i\cdot t^i$, 
where $i\in\Z$ 
and 
where $a_i\neq0$ holds for finite numbers of $i$. 
Take a closed oriented 3-manifold $M_i$ such that $\tau_7(M_i)=a_i$ 
(of course, 
$\tau_7(\quad)$ can be changed into another appropriate invariant). 
Then, anyway, $\{M_i\}$ is a parallel world for the Jones polynomial. 
Of course, this way is very trivial. 
We want a way to make parallel world that we can make it 
without information of $V_K(t)$. 
We need to find a `good' parallel world. 
But the condition `without information of $V_K(t)$' is very vague 
and the word `good' is used in the meaning in daily life. 
So, we would say that 
the concept of parallel world for observable 
is principle to find a new TQFT.

If an invariant for manifolds has a good vacuum parallel world, 
then we only have to investigate a Lagrangian. 
We do not need to 
investigate both Lagrangian and observable spontaneously. 
In this meaning, the concept of parallel world is useful in order 
to find a new invariant.

For complex manifolds, 
we know a Lagrangean whose path integral we can compute, 
e.g. holomorphic Chern-Simons term 
(see 
\cite{HollowoodIqbalVafa} 
\cite{Thomas}
\cite{WittenCSstring}
and their references). 
We know topological modification of complex manifolds 
compatible with their complex structures, e,g., blow up, flop.  
What is obdervables corresponding to these modifications?

For high dimensional knots 
(see \cite{CochranOrr}
\cite{LevineOrr}
\cite{Ogasa} 
\cite{Rolfsen}
and their references.),   
we know a classical Lagrangean, e.g. Chern-Simons $n$-form 
(note $n$ is odd; see \cite{ChernSimons}), 
which would be able to be quantized. 
We know topological modification in high dimensional knot theory, 
using surgery, branched cyclic covering, complement etc.
Can we find a new model of TQFT in high dimensional knot theory?


\footnotesize{
 }
\end{document}